\documentclass{article}
\usepackage{spconf,amsmath,graphicx}

\usepackage{url}
\usepackage[hidelinks]{hyperref}
\usepackage{bm}
\newcommand{\norm}[1]{\left\lVert#1\right\rVert}
\usepackage[font=small]{caption}
\usepackage[font=footnotesize]{subcaption}

\title{Multitask Learning with Capsule Networks for Speech-to-Intent Applications}
\name{Jakob Poncelet \thanks{This research received funding from the Flemish Government under the
“Onderzoeksprogramma Artifici\"ele Intelligentie (AI) Vlaanderen”
programme.}, Hugo Van hamme}
\address{KU Leuven \\ 
     Department Electrical Engineering ESAT-PSI\\
     Kasteelpark Arenberg 10, Bus 2441, B-3001 Leuven Belgium\\
     \texttt{\{jakob.poncelet, hugo.vanhamme\}@esat.kuleuven.be}
      }
\begin{document}
%
\maketitle
\begin{abstract}
Voice controlled applications can be a great aid to society, especially for physically challenged people. However this requires robustness to all kinds of variations in speech. A spoken language understanding system that learns from interaction with and demonstrations from the user, allows the use of such a system in different settings and for different types of speech, even for deviant or impaired speech, while also allowing the user to choose a phrasing. The user gives a command and enters its intent through an interface, after which the model learns to map the speech directly to the right action. Since the effort of the user should be as low as possible, capsule networks have drawn interest due to potentially needing little training data compared to deeper neural networks. In this paper, we show how capsules can incorporate multitask learning, which often can improve the performance of a model when the task is difficult. The basic capsule network will be expanded with a regularisation to create more structure in its output: it learns to identify the speaker of the utterance by forcing the required information into the capsule vectors. To this end we move from a speaker dependent to a speaker independent setting.
\end{abstract}
\begin{keywords}
Spoken Language Understanding, Capsule Networks, Multitask Learning, End-to-end, Speaker Identification
\end{keywords}
\section{Introduction}
\label{sec:intro}

Technology is advancing at an unprecedented rate, ultimately trying to ease and improve the life of people. Speech recognition is playing a major role in this trend to allow hands-free operation of all kinds of devices. Voice control using spoken language understanding (SLU) systems can be beneficial in all parts of daily life, but more specifically it would help physically challenged and elderly people to live independently. Command-and-control (C\&C) applications are typical in this setting, e.g. for positioning of their bed, operating domestic devices, etc. However this requires the system to understand non-standard speech as well, like thick dialects or impaired speech, which is more frequent in these user groups. This is where common speech technology based on acoustic models runs into problems \cite{ASR}. A speech-to-intent understanding system can be more robust to variations and errors in speech, since it doesn't use an intermediate textual representation, and is attracting more and more research interest \cite{SLU2, SLU3, SLU4}.

In \cite{vrenkens_assist} an SLU system for C\&C has been implemented, which builds up a model from scratch using demonstrations from the user. The system learns to map the spoken commands uttered by the user directly to a semantical representation with labels for every task (speech-to-intent). Building a model from scratch from user demonstrations, i.e. without making linguistic assumptions such as the phone set, vocabulary or grammar, makes it also accessible for deviant speech and multiple application and language domains. Moreover it allows the user to choose how to phrase the commands, instead of being confined to the wording chosen by the designer.

In this paper the implementation of the aforementioned SLU system, which is built with capsule networks, will be analysed and adapted. Capsule networks were presented in 2017 by Hinton \cite{capsule1} and are a new type of deep neural network (DNN), believed to need less training data than standard DNN's. The proposed capsule networks have been compared in accuracy and data requirements in this setting to a previously proposed Non-negative Matrix Factorisation (NMF) approach \cite{aladin1,aladin2}. The capsule network was deemed very promising, since it often outperformed the other architectures \cite{capsule4}, hence an insight into its working would be useful.

A capsule network consists of layers of capsules, with each capsule being characterised by a vector (as compared to scalar neurons). The activation vectors of the capsules in the output layer are essentially a condensed representation of the information that the network uses to classify the speech to the right intent. The effect of the dimension of these vectors is examined to get an indication of its importance and the benefit of giving the capsules more freedom for the orientation of the activation vectors. If a low-dimensional space would suffice without introducing errors, the number of parameters can be reduced to improve efficiency.

Besides allowing the capsules to freely put in their output whatever they like for the classification of an utterance, one can also force them to use the dimensions of the vectors to store information and create a more structured output. The network can be extended to learn auxiliary tasks at the same time because of these higher dimensional vectors. In other words, multitask learning can be implemented by applying regularisations to the output vectors, which could improve the efficiency and performance of the model \cite{multitasklearning, multitask2}. In this paper, the model will learn to identify the speaker that uttered the command by encapsulating information into the orientation of the capsule vectors. Learning which speaker gave the command is a useful task for the system to incorporate when decoding the utterances and might improve the performance \cite{speakers}. Consequently we move from a speaker dependent setting (as is \cite{aladin2}) to a speaker independent setting. Training a model with mixed data from multiple speakers gives a penalty in learning speed, since different speakers use different phrasings for their commands and acoustic speaker variation needs to be learnt as well. From a practical point of view, speaker identification allows an SLU system to be shared by multiple users and the system would be able to independently figure out for which person a task has to be carried out (which could be different from person to person).

The basic and extended model will first be explained in section 2, along with some basic theory about capsule networks. Section 3 discusses specifics about the methods used in the experiments and section 4 describes the performed experiments. In section 5 the results are discussed and section 6 finally gives a conclusion to this work.

\section{Model}
\label{sec:model}
\subsection{Capsule Network Baseline Model}
\label{ssec:capsnet}
A capsule network consists of different layers of capsules. A capsule is characterised by an activation vector $\bm{u}_i$. The length of this vector corresponds to the probability of an object being present, and the orientation of this vector corresponds to the parameters of the object (for example the pose). Every capsule in a layer will try to predict the output of the capsules in the next layer. This prediction uses a transformation matrix $\bm{W}_{ij}$ for every capsule pair in consecutive layers that will be learned by backpropagation of the loss through the network.
\small
$$
\hat{\bm{u}}_{j|i} = \bm{W}_{ij}\bm{u}_i \eqno{(1)}
$$
\normalsize
The connection between two layers uses a dynamic routing algorithm, as explained in \cite{capsule1}, and is based on agreement between predictions of the high level capsule property by the lower level capsules. After a few iterations the output vector of the capsules in the subsequent layer is obtained. Note that the length of this vector is between 0 and 1 using a squash function. Finally for classification purposes a margin loss can be implemented proceeding from the length of the activation vectors $\bm{v}_k$ of the last capsule layer. For $K$ classes this is defined as in (2), with $T_k$ equal to 1 if class $k$ is present (and 0 otherwise), and $m^{+}=0.9$ and $m^{-}=0.1$ chosen boundary values. Every output capsule corresponds to a task label and the decoded task is decided based on the most active capsules, i.e. with an activation vector with norm close to 1
\small
$$
L_l = \sum_{k=1}^{K} T_k \: max(0,m^{+}-\norm{\bm{v}_k}) + (1-T_k) \: max(0,\norm{\bm{v}_k}-m^{-}) \eqno{(2)}
$$
\normalsize
A more detailed explanation about capsule networks can be found in \cite{capsule1}. The implementation in \cite{vrenkens_assist,capsule4} with two layers (a primary capsule layer and an output capsule layer) serves as the baseline model that will be used for the experiments in this paper.

\begin{figure*}
    \centering
    \includegraphics[scale=0.64]{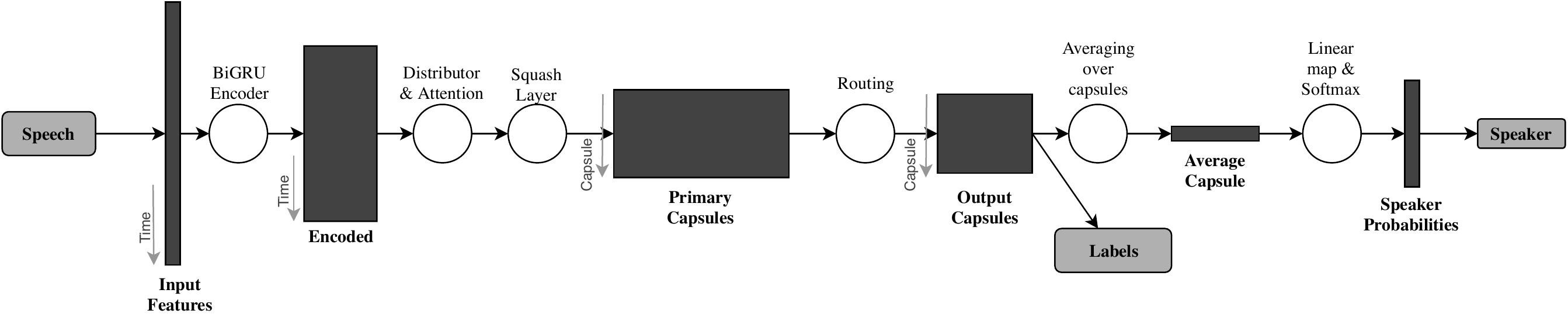}
    \caption{Schematic of the multitask model}
    \label{fig:diagram}
\end{figure*}

\subsection{Multitasking Model with Speaker Identification}
\label{ssec:expanded}
We want to give a meaning to the dimensions of the output capsules through multitask learning, so the model will use them more actively and create a more structured output. This was previously not explicitly required of the capsules, since the classification is only based on the length of the output vectors (as in (2)), not on the orientation. In this section the baseline model is extended with an additional layer to learn which speaker spoke the command. 

We start with a definition of the average capsule $\bm{z}$ in each utterance, with $N$ the number of output capsules.
\small
$$
    \bm{z} = \frac{\sum\limits_{i=1}^{N} \bm{v}_i}{\sum\limits_{i=1}^{N} \norm{\bm{v}_i}} \eqno{(3)}
$$
\normalsize
The average capsule combines for every output dimension the information of the vectors of all output capsules (it averages over them). The average capsule is thus a column vector of dimension equal to the dimension of the output capsules, for example 8. A single-layer neural network followed by a softmax layer will map the average capsules to speaker probabilities. The weight matrix of this layer will be called the projection matrix $\bm{W_s}$ and has dimensions $\left( n \times M \right)$, with $n$ the output dimension of the capsules and $M$ the number of speakers in the dataset. In the testing phase, the model chooses the speaker with the highest probability. A schematic of the multitask model is shown on Fig.~\ref{fig:diagram}.

To let the model learn, a new speaker loss term is added to the total loss (for now only consisting of the label loss as in (2)). The speaker loss uses a cross-entropy loss function based on the target speaker (as a one-hot encoded vector $\bm{t}$) and the estimated probabilities $P_i$ for every speaker $i$, and is summed over all $M$ speakers.
\small
$$
    L_{s} = -\sum_{i=1}^{M} t_{i} log(P_i) \eqno{(4)}
$$
\normalsize
Backpropagation of the loss will then adapt the trainable matrix $\bm{W_s}$ to correctly identify the speaker and will force the capsules to encapsulate this into their orientation. A regularisation parameter $\lambda_s$, defined as the speaker weight, is added in the total loss function to weigh the relevance of the speaker loss compared to the label loss.
\small
$$
    L_{tot} = L_l + \lambda_{s}L_s \eqno{(5)}
$$
\normalsize

\section{METHOD}
\subsection{Dataset}
The model will be tested on two publicly available datasets. The GRABO dataset \cite{Grabo,GraboPaper} is based on a setting where a person gives commands to a robot. The robot can move around, pick things up and point a laser. There are a total of 33 different output labels corresponding to possible positions, movement speeds and actions. Data has been recorded from ten Dutch speakers and one English speaker. With around 6000 recorded utterances, this is a smaller dataset with little variety, e.g. most commands have the same structure of sentences.

The Fluent Speech Commands dataset by Fluent.ai \cite{Fluent,Fluentpaper} is a larger and more challenging dataset. It comprises 30000 utterances from 97 speakers, used in a smart-home controlling appliance setting, for e.g. controlling the lights or music volume in a certain room. There are 31 unique intent labels, but there is much more variation in the spoken commands. We should point out that there are some speakers with only a few recorded utterances.

\subsection{Experimental Setup}
Most of the experiments in this paper are cross-validation experiments. The dataset is divided into 150 blocks. Starting from one block, the model will be trained on an increasing number of blocks, and tested on all remaining blocks. This way a learning curve is created. In the speaker dependent experiments with the baseline model, the data is fed to the model speaker by speaker and the final curve is obtained by averaging over the results for every speaker. On the contrary, in the (speaker independent) experiments involving speaker identification, the utterances from all speakers are randomly shuffled beforehand and then all data is divided into blocks. 

The hyperparameters of the model are chosen as in \cite{vrenkens_assist} and are not altered, except when specifically mentioned. The varying parameters will be the dimension of the output capsules and the regularisation weight of the multitask model.

Evaluation of the label classification task is done using an F1 score \cite{F1}. To evaluate the identification of the speakers, we use the percentage of correctly decoded speakers.

\section{EXPERIMENTS}
First of all the effect of the dimension of the output capsules was analysed by comparing experiments for different output dimensions (ranging from 2 to 8). The analysis was done with the baseline model on the GRABO dataset in a speaker dependent setting. We observed that the output capsule dimension had little to no impact on the F1 scores, even down to a capsule dimension of 2.

\begin{figure}[htpb]
    \centering
    \subcaptionbox{\label{fig:speaker1}}{\includegraphics[width=0.49\linewidth]{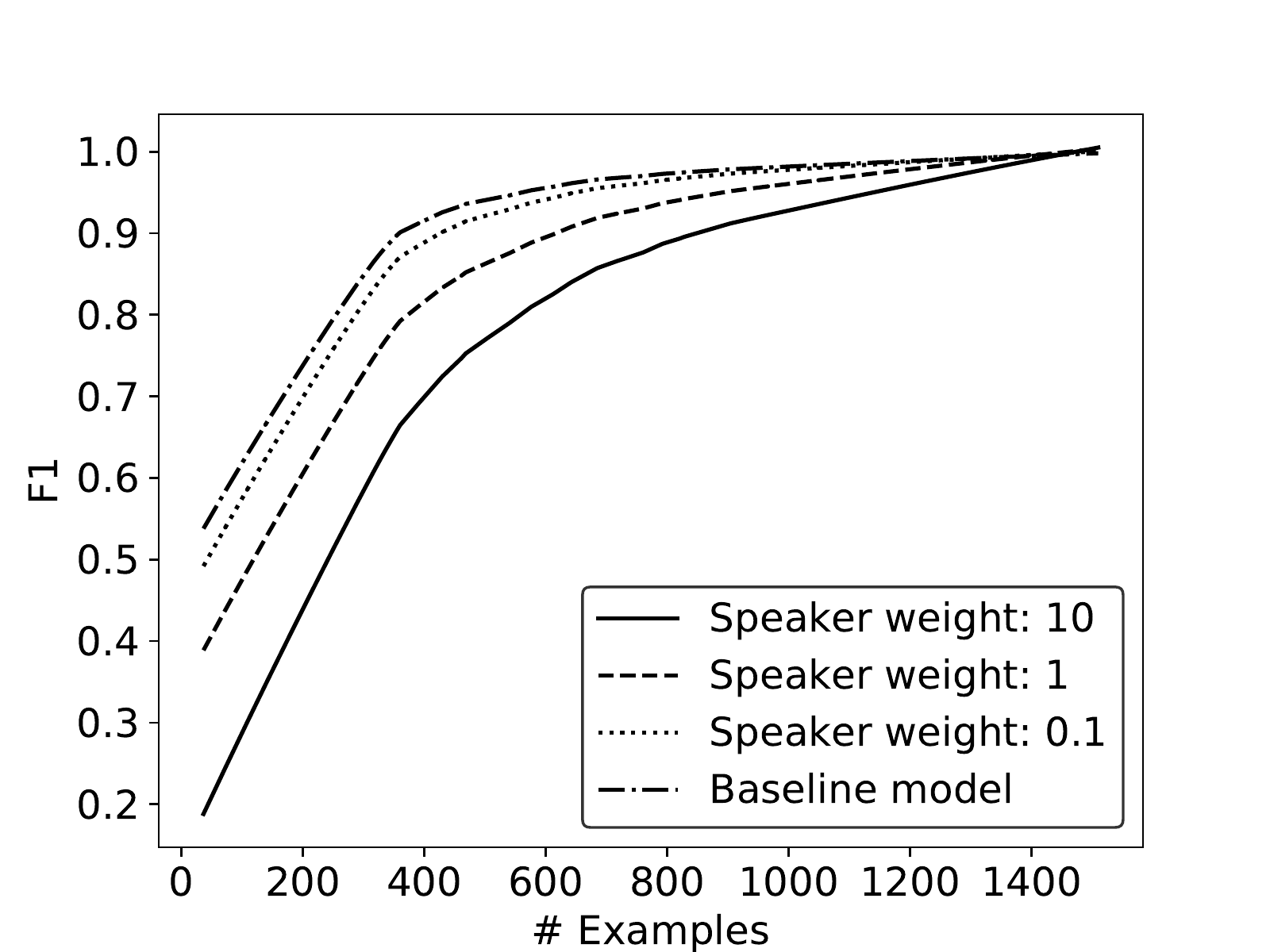}}
    \subcaptionbox{\label{fig:speaker2}}{\includegraphics[width=0.49\linewidth]{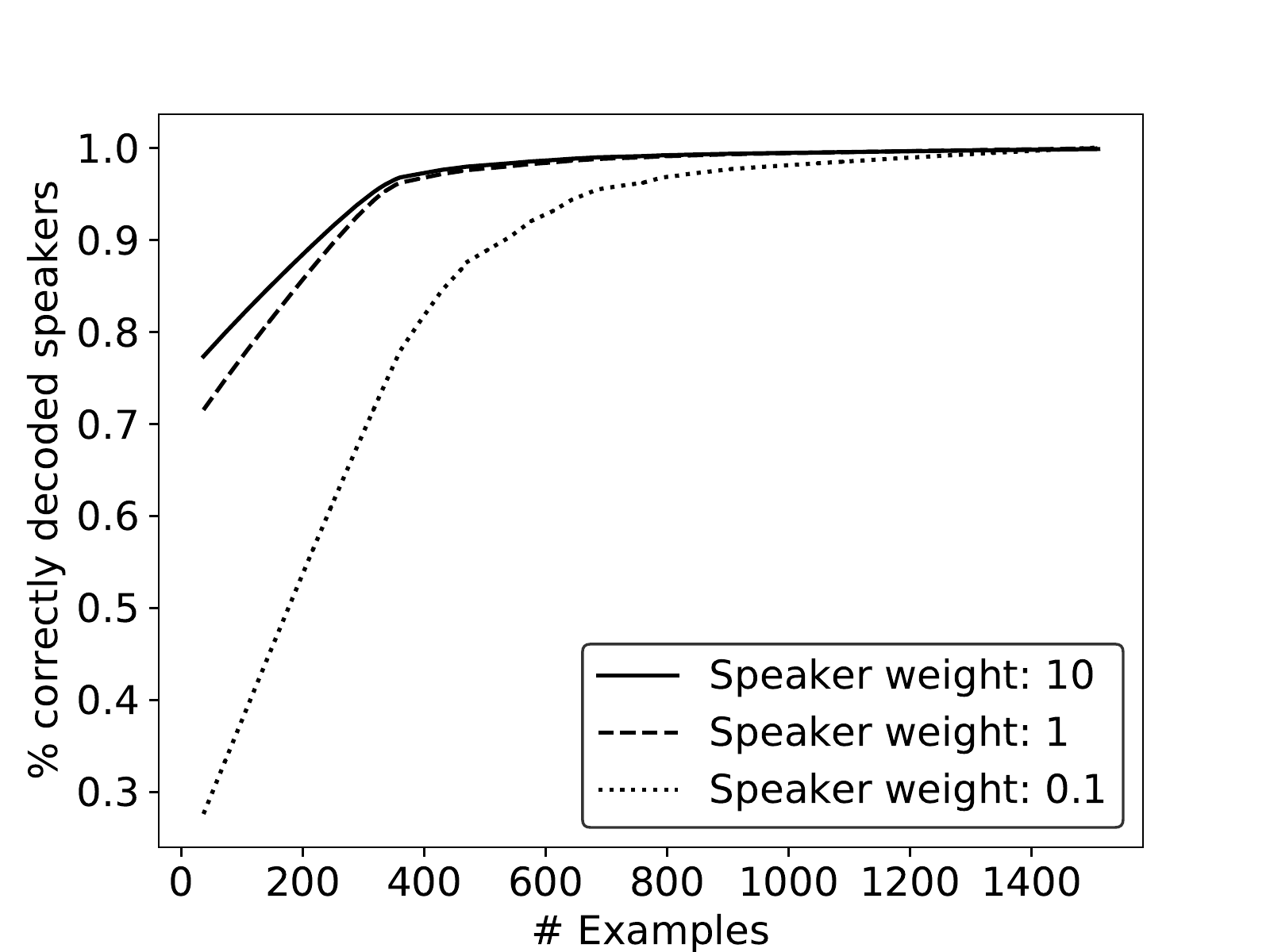}}
    \subcaptionbox{\label{fig:speaker3}}{\includegraphics[width=0.49\linewidth]{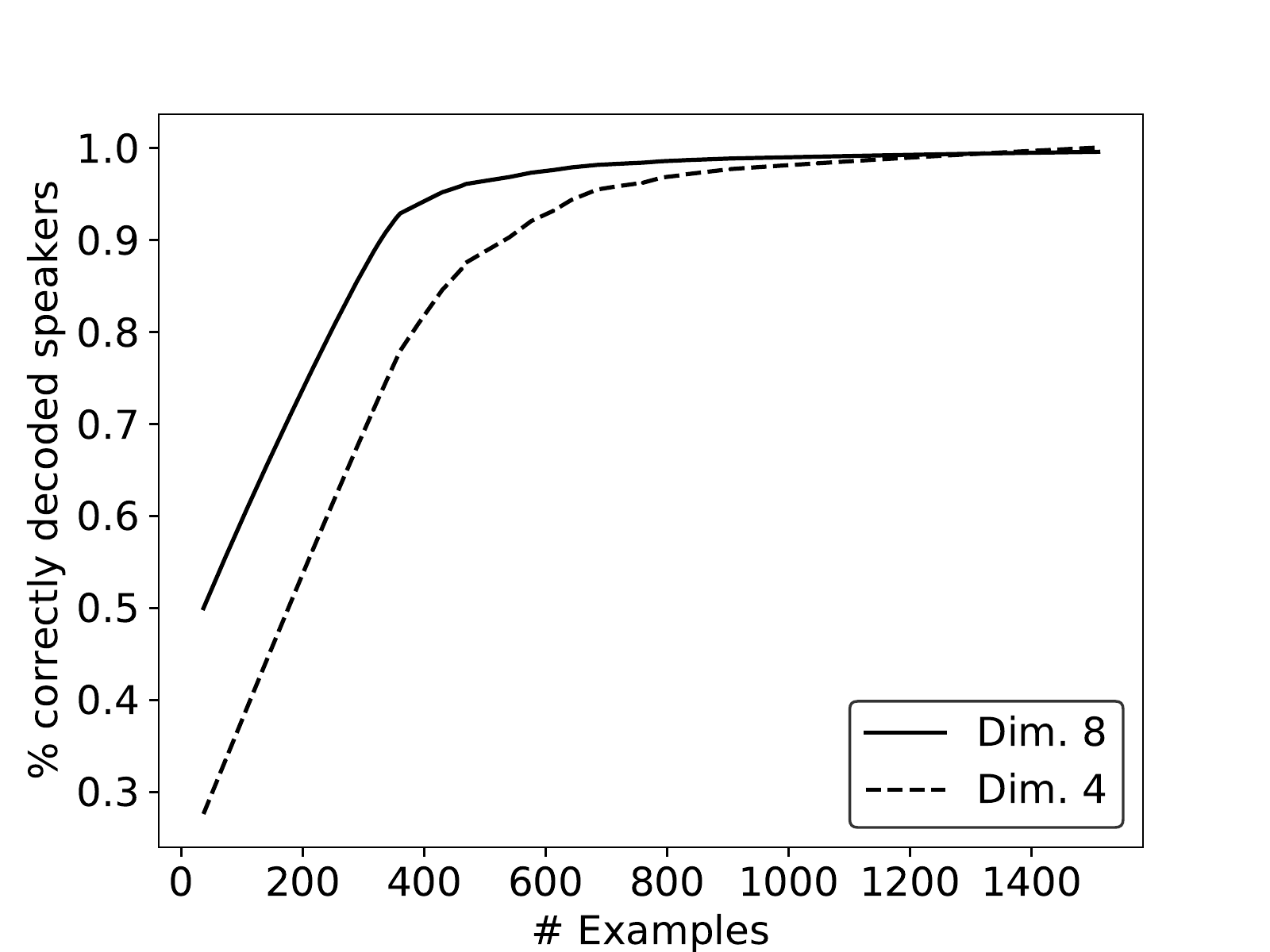}}
    \caption{Experiments with the multitask model on GRABO for different parameters, showing the effect of the speaker weight on the F1 score in (a) and on the speaker identification capability in (b), and the effect of the output dimension on speaker identification in (c).}
    \label{fig:speakers}
\end{figure}

\noindent Afterwards the multitask model with speaker identification was analysed. Multiple speaker independent experiments have been executed on GRABO, first with different factors for the speaker weight $\lambda_s$ (and a fixed output dimension of 4) and second with different output dimensions. Fig. \ref{fig:speaker1} shows the effect of the added speaker loss term on the F1 score, compared to the performance of the baseline model without speaker identification. Fig. \ref{fig:speaker2} compares the speaker recognition of experiments with speaker weights 10, 1 and 0.1. Fig. \ref{fig:speaker3} shows the result of experiments with output dimension 4 and 8 and a speaker weight of 0.1. The F1 score was the same for both experiments and is thus not shown.

Fig. \ref{fig:FS_result} shows the results of cross-validation experiments performed on the Fluent Speech Commands dataset, comparing the multitask model to the baseline in a speaker independent setting. The speaker weight regularisation parameter of the multitask model has been set to 1 and the output dimension of the capsules to 16. 

Finally the train and test experiments of \cite{Fluentpaper} for the Fluent Speech Commands dataset have been replicated for comparison. Using the accuracy metric as defined in that paper, the multitask model achieved an accuracy of 97.8\% on the test set after training on the partial dataset and 98.1\% after training on the full dataset. These results should be compared to the model without pre-training of \cite{Fluentpaper}, which reaches an accuracy of 88.9\% with the partial dataset and 96.6\% with the full dataset.

\section{DISCUSSION}
From the first analysis we conclude that the output capsule dimension can be reduced without introducing errors to lower the number of parameters. The network almost solely uses the length of the vectors. If there is information in the orientation of the vector, it can be presented in two dimensions, so there is probably not much structure present in the capsules.

Fig.~\ref{fig:speaker2} confirms that the speakers are successfully identified in the multitask model. The performance is already at 99\% after a few hundred examples, which means this task is not so difficult for the model, since there are only 11 speakers in the GRABO dataset. There is a trade-off between the speaker identification and the task learning speed, depending on the speaker weight $\lambda_s$, as presented on Fig.~\ref{fig:speaker1}. This negative effect on the F1 score can be explained by the fact that the task of extracting the labels is quite easy for the model on the GRABO dataset, due to the little variability in phrasings (the F1 score reaches over 90\% after a few hundred examples). However based on the comparison between different dimensions of Fig.~\ref{fig:speaker3}, we see that now the output dimension does have an influence and the orientation of the output vector has received more meaning compared to the baseline model.

\begin{figure}[ht]
    \centering
    \subcaptionbox{ \label{fig:FS_1}}{\includegraphics[width=0.49\linewidth]{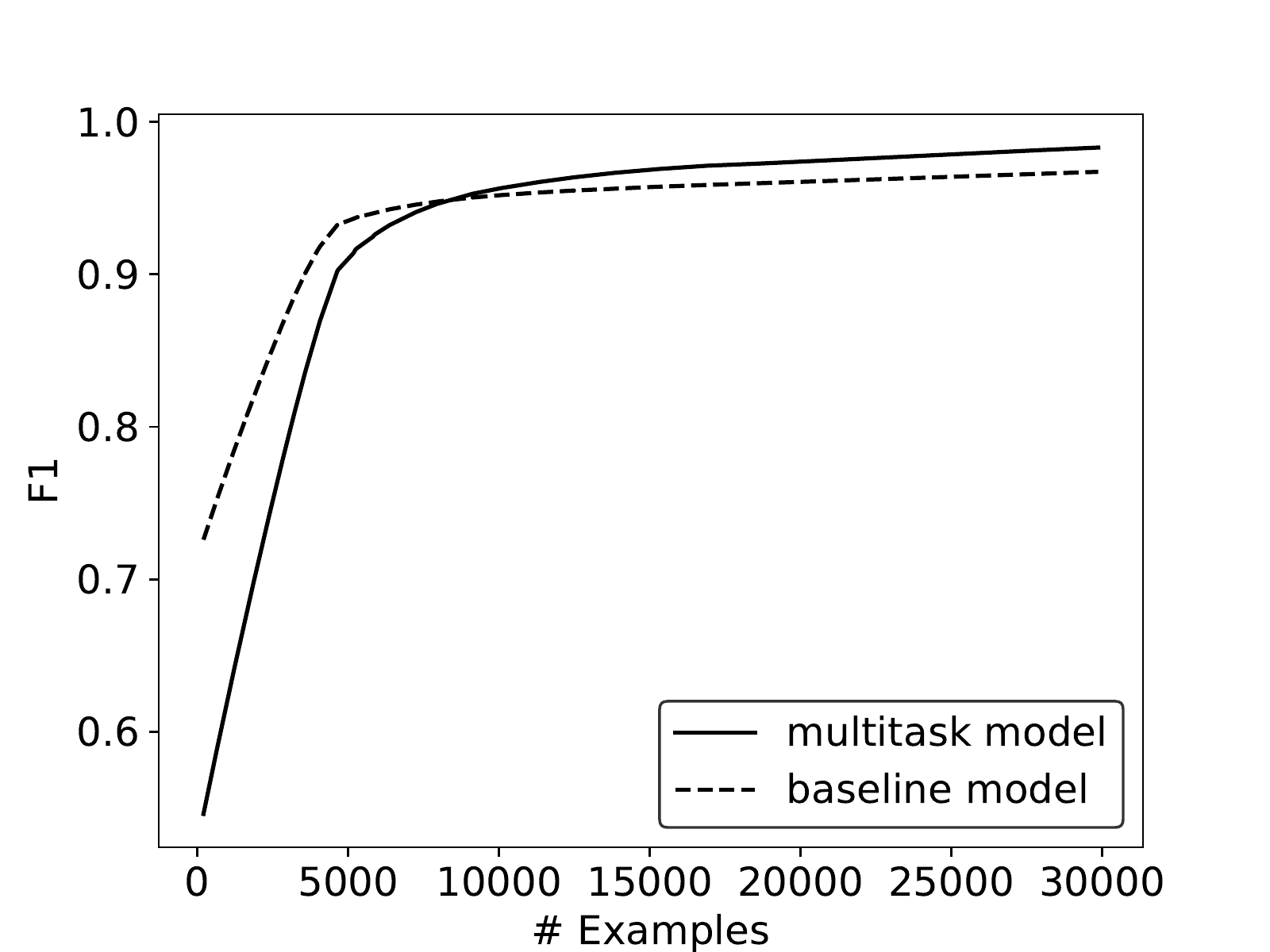}}
    \subcaptionbox{ \label{fig:FS_2}}{\includegraphics[width=0.49\linewidth]{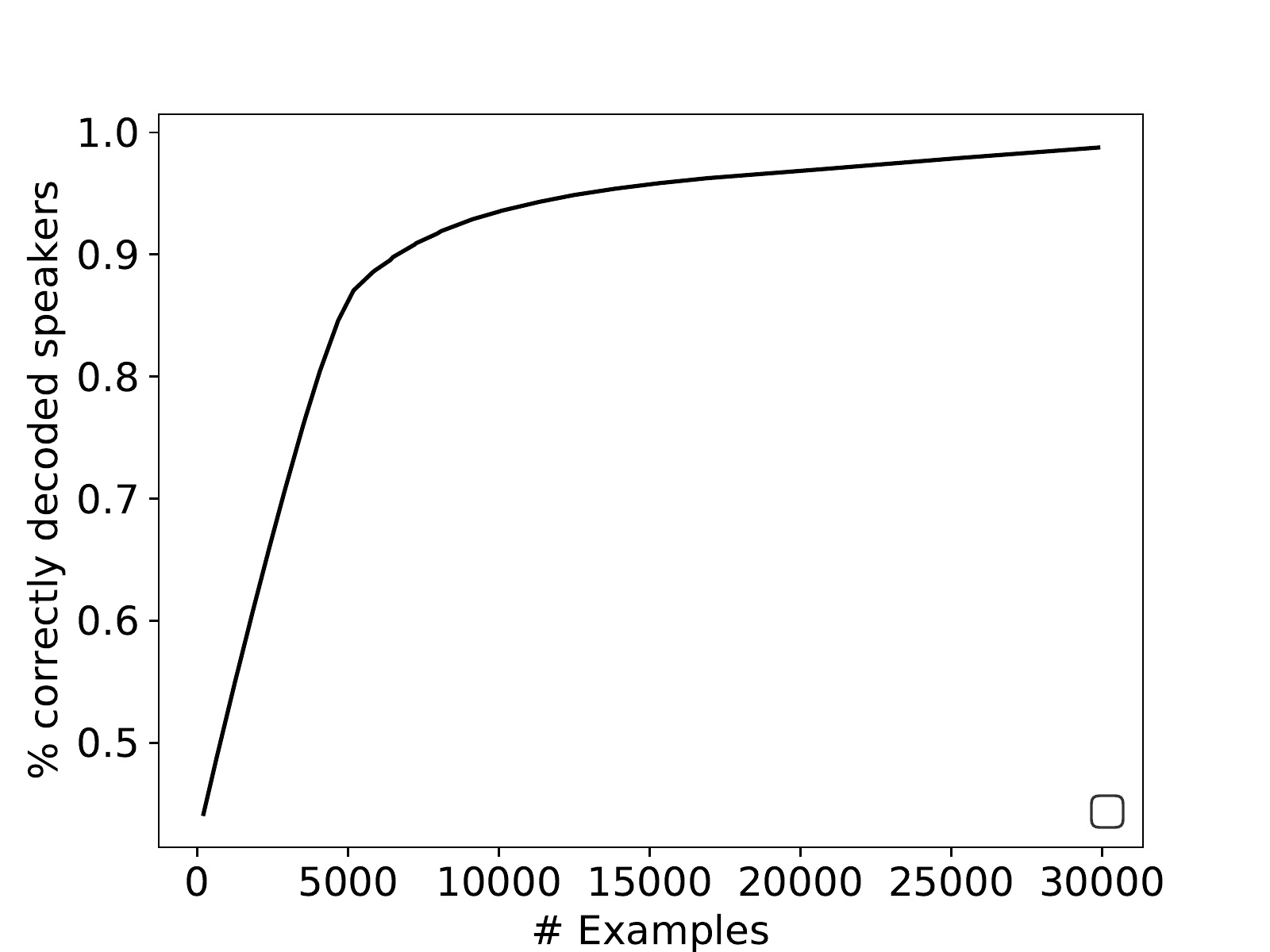}}
    \caption{Experiments with the multitask model on the Fluent Speech Commands dataset, comparing the F1 score to the baseline model in (a) and showing the speaker recognition performance for a speaker weight of 1 in (b).}
    \label{fig:FS_result}
\end{figure}

Fig.~\ref{fig:FS_result} shows that on the larger, more difficult Fluent Speech Commands dataset, multitask learning has improved the asymptotical performance. The learning speed is slower in the multitask model (the performance is worse when little training data is available), because the added term in the loss function will make the model initially less focused on the decoding task. Once the speakers are reliably recognised, this will help the intent decoding. With nearly 100 speakers in the dataset, the model needs enough examples to be able to make a distinction between all those speakers to identify the right one. Finally the accuracy results of the multitask model on the train and test experiment in \cite{Fluentpaper} are higher than the results of the model (without pre-training) proposed there.

\vfill
\section{CONCLUSION}
In this paper we investigated the use of capsule networks as fast learning models for speech-to-intent systems, or more specifically for command-and-control applications. Analysis of the basic capsule network showed that there was not much information encapsulated in the orientation of the output vector. The length of the vector is most important for the classification task.

The baseline model has been expanded to incorporate multitask learning in the capsule vectors and in order to create more structure in its output. We moved to a speaker independent setting, and the model now also learns to identify the speaker of the utterance. For this auxiliary task a linear mapping is introduced on the average output capsules to combine their dimensions and use them for learning. From the results we can conclude that this regularisation has led to structure in the output capsule, reflecting speaker identity. Furthermore identifying the speaker and encoding the required information structurally into the orientation of the capsule vectors has improved the performance of the model when the dataset is challenging and large enough. It is remarkable to see that our model performs well even on the Fluent Speech Commands dataset, where there are some speakers with only very few recorded utterances.

%

\vfill

\raggedright

\bibliographystyle{IEEEbib}
\bibliography{references}

\end{document}